\documentstyle[pre,multicol,aps]{revtex}

\begin{document}

\setlength{\textwidth}{180mm}
\setlength{\textheight}{240mm}
\setlength{\parskip}{2mm}

\input{epsf.tex}

\title{ Parametric localized modes in quadratic nonlinear 
        photonic structures }
\author{Andrey A. Sukhorukov$^{1}$, Yuri S. Kivshar$^1$, 
        Ole Bang$^{1,2}$, and Costas M. Soukoulis$^3$ }

\address{
$^1$ 
Optical Sciences Centre, Australian National University, Canberra ACT 0200, Australia
\\
$^2$  Department of Mathematical Modelling, Technical University of Denmark, DK-2800 Lyngby, Denmark
\\
$^3$ Ames Laboratory and Department of Physics and Astronomy, Iowa State University, Ames, Iowa 50011
}

\maketitle

\begin{abstract}
We analyze two-color spatially localized modes formed by parametrically coupled fundamental and second-harmonic fields excited at quadratic (or $\chi^{(2)}$) nonlinear interfaces embedded into a linear layered structure~--- a {\em quasi-one-dimensional quadratic nonlinear photonic crystal}. For a periodic lattice of nonlinear interfaces, we derive an effective discrete model for the amplitudes of the fundamental and second-harmonic waves at the interfaces (the so-called discrete $\chi^{(2)}$ equations), and find, numerically and analytically, the spatially localized solutions~--- {\em discrete gap solitons}. For a single nonlinear interface in a linear superlattice, we study the properties of two-color localized modes, and describe both similarities and differences with quadratic solitons in homogeneous media.
\end{abstract}


\vspace*{-1.0 cm}

\begin{multicols}{2}

\narrowtext

\section{Introduction}

The physics and applications of photonic band gap materials (or {\em photonic crystals}) have been an active topic of research for more than a decade~\cite{book}. The next step in application of photonic crystals is to create {\em tunable band gap materials} where the gap could be controlled by an external parameter. One of the recent suggestions~\cite{recent1} is based on the inverse opal structure~--- a microscopic lattice of spheres of air surrounded by silicon~--- with a layer of liquid crystal material that can make the transmission properties programmable by applying an electric field. Continuous tuning can also be realized by varying the liquid crystal temperature~\cite{recent2}. 

Another important idea towards creating {\em dynamically tunable} band gap materials for switches and transistors operating entirely with light is to employ their nonlinear properties, thus creating {\em nonlinear photonic crystals}. The concept of nonlinear photonic crystals, defined as having a spatially periodic nonlinearity, was introduced by Scalora {\em et. al.} \cite{Scalora94} in a numerical study of ultra-fast optical switching and limiting in cubic nonlinear Kerr materials. Bistability and localized modes in photonic superlattices with embedded layers possessing nonresonant cubic (or Kerr-type) nonlinearities have also been discussed in the literature~\cite{souk}. Recent advances in the so-called {\em cascaded nonlinearities}~\cite{chi2} demonstrate an effective way to lower the switching power by employing parametric interaction and frequency conversion in noncentrosymmetric quadratic nonlinear optical materials. Parametric interactions are also known to support solitary waves, {\em spatial quadratic solitons}~\cite{solitons}, that exist in homogeneous media where spatial localization is induced by two-wave parametric mixing processes between the fundamental wave and its second harmonic. Quadratic nonlinear photonic crystal was introduced as a concept by Berger~\cite{nonPC} in a study of multi-wavelength frequency conversion.

Taking into account the similarities between the localized defect modes in linear inhomogeneous media and nonlinear localized modes in homogeneous media~\cite{braun}, we wonder if parametric interactions can support localized modes in a variety of photonic band gap structures with quadratic (or $\chi^{(2)}$) nonlinearities. The first step in this direction has been recently presented in Ref.~\cite{PRE}, where we have analyzed second-harmonic generation (SHG) at a thin, effectively quadratic nonlinear layer separating two (generally different) homogeneous linear media, and predicted multistability of SHG for both plane waves and localized modes, also describing two-color localized photonic modes that can be excited at the interface.

The main purpose of this paper is twofold. First, we generalize the results of Ref.~\cite{PRE} to the case of a thin quadratic nonlinear layer embedded into an arbitrarily stratified periodic linear medium. In particular, we consider a nonlinear defect layer with second-order nonlinear response in a perfectly periodic dielectric structure~--- a one-dimensional analog of a photonic crystal with a nonlinear impurity. Secondly, we develop a general formalism for analyzing localized modes in multilayer structures~-- 
{\em nonlinear superlattices}~--- and describe {\em two-color localized gap modes} supported by a periodic lattice of thin layers with quadratic nonlinearity.

The structure of the paper is the following. In Sec.~\ref{sect:model} we present our model that is described by a system of two coupled nonlinear equations for the envelopes of the fundamental and second-harmonic waves. In the stationary case, the solution can be presented as a superposition of the forward and backward traveling waves, and this allows to derive an effective system of discrete coupled-mode equations for the wave amplitudes at the layers, the so-called discrete $\chi^{(2)}$ equations (Sec.~\ref{sect:period_general}). Solutions of these equations for localized nonlinear modes are briefly discussed in Sec.~\ref{sect:period_dgsol}. At last, in Sec.~\ref{sect:single} we consider a generalization of the results of Ref.~\cite{PRE} to the case of a single quadratic nonlinear layer embedded into a periodic linear medium, and describe a number of new features of the localized modes that appear due to a bandgap structure.

\section{Model} \label{sect:model}

First, we discuss the physical motivation for our model. Let us consider an interface between two semi-infinite bulk optical media with inversion symmetry. The interface layer breaks the symmetry and therefore it should possess a nonvanishing surface quadratic response~\cite{agran} that can be enhanced by a proper coating creating a nonlinear layer with quadratic nonlinearity~\cite{vilaseca2}. Such a layer corresponds to an effective nonlinear defect that can support photonic modes localized at the interface. There exists a strong experimental evidence of SHG in such type of localized photonic modes. For example, recent experimental results~\cite{vilaseca} reported SHG in periodic photonic band-gap structures with embedded nonlinear defect layers. An enhancement of the parametric interaction in the vicinity of the defects was observed, suggesting that SHG occurs in localized modes, being suppressed for propagating modes. The surface nature of SHG was confirmed by comparison of experimental data and results of direct numerical simulations~\cite{vilaseca2}.

\begin{figure}
\setlength{\epsfxsize}{11cm}
\vspace*{-7mm}
\hspace*{-0.8cm}\centerline{\mbox{\epsfbox{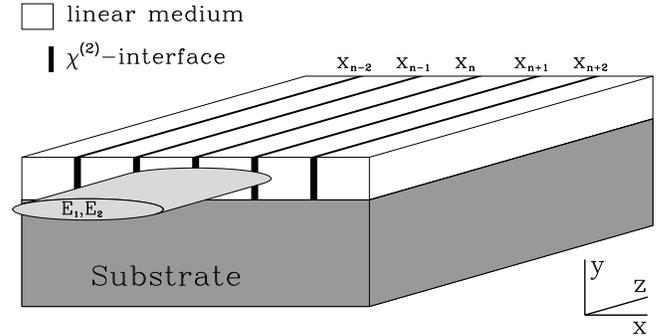}}}
\vspace*{-7mm}
\caption{ \label{fig:model}
Schematic diagram of an array of thin $\chi^{(2)}$ layers
embedded into a linear slab waveguide, with indication of
the direction of the input laser beam relative to the interfaces.
}
\end{figure}

To introduce an analytically solvable model for SHG in localized waves, we follow Ref.~\cite{PRE} and consider a fundamental frequency (FF) wave propagating along the $Z$-direction in a linear slab waveguide, as shown in Fig.~\ref{fig:model}. We assume that the interfaces (or {\em defect layers}) possess a quadratic nonlinear response, so that a FF wave can parametrically couple to its second harmonic (SH) wave. The coupled-mode equations for the complex envelope functions $E_j(x,Z)$ (we use $j$=1,2 for FF and SH, respectively) can be written in the form,
\begin{equation} \label{eq:chi2N} 
\begin{array}{l} 
  {\displaystyle  i \frac{\partial E_1}{\partial Z} + 
       D_1 \frac{\partial^2 E_1}{\partial x^2} 
         + \varepsilon_1(x) E_1 + \Gamma_1(x) E_1^* E_2 = 0, } 
                   \\*[9pt]
  {\displaystyle i \frac{\partial E_2}{\partial Z} + 
       D_2 \frac{\partial^2 E_2}{\partial x^2} 
         + \varepsilon_2(x) E_2 + \Gamma_2(x) E_1^2 = 0,}
   \end{array}
\end{equation}
where $D_j$ are the diffraction coefficients ($D_j>0$). In the approximation of infinitely thin interface layers (valid when the width of each layer is much smaller than the characteristic transverse scale of the FF and SH wave envelopes), we take $\varepsilon_j(x) = \varepsilon_{0j}(x) + \sum_n \kappa_j \delta( x - x_n)$ and $\Gamma_j(x) = \sum_n \gamma_j \delta(x - x_n)$, where $x_n$ denotes the position of the \mbox{$n$-th} nonlinear interface, $\gamma_j$ are the nonlinearity coefficients, $\varepsilon_{0j}(x)$ and $\kappa_j$ account for the phase velocity differences in bulk and interface materials. 

In order to reduce the number of physical parameters, we normalize 
Eqs.~(\ref{eq:chi2N}) as follows: 
$E_1(Z)=u(z) e^{i \; \overline{\varepsilon}_{01} Z} /\sqrt{\gamma_1\gamma_2}$, 
$E_2(Z)=v(z) e^{2 i \; \overline{\varepsilon}_{01} Z}/\gamma_1$, 
$\sigma=D_2/D_1$, 
$\nu_j(x) = (\varepsilon_{0j}(x) - j \; \overline{\varepsilon}_{01}) / D_1$, 
$\overline{\varepsilon}_{0j}$ is the average value of 
$\varepsilon_{0j}(x)$, and $\beta_j = \kappa_j/D_1$,
where $z=Z/D_1$ is measured in units of $D_1$. 
Then, the dimensionless equations become
\begin{equation} \label{eq:chi2}  
  \begin{array}{l} 
   {\displaystyle 
      i \frac{\partial u}{\partial z}
      + \frac{\partial^2 u}{\partial x^2} + \nu_1 \left( x \right) u
   } \\*[9pt] {\displaystyle \qquad
       + \sum_n \delta \left( x - x_n \right)
         \left( \beta_1 u + u^* v \right) = 0 , 
   } \\*[9pt] {\displaystyle 
      i \frac{\partial v}{\partial z}
      + \sigma \frac{\partial^2 v}{\partial x^2}
      + \nu_2 \left( x \right) v 
   } \\*[9pt] {\displaystyle \qquad
       + \sum_n \delta \left( x - x_n \right)
         \left( \beta_2 v + u^2 \right) = 0 .
   } \end{array}             
\end{equation}
At this point, it is important to note that our equations 
[Eqs.~(\ref{eq:chi2N}) or (\ref{eq:chi2})] describe the beam evolution 
in the framework of the so-called {\em parabolic approximation}, 
valid for the rays propagating mainly along the $Z$ direction. 
In other words, the characteristic length of the beam distortion due to
diffraction and refraction along the $Z$ axis should be much 
larger than the beam width in the transverse direction $x$.
This leads to the requirement of {\em a shallow grating}, 
$|\varepsilon_{0j}(x) - \overline{\varepsilon}_{0j}| 
                                  \ll \overline{\varepsilon}_{0j}$. 
On the other hand,
to make the parametric interaction effective, the mismatch between 
the phase velocities of the FF and SH waves should be small, 
$2 \overline{\varepsilon}_{01} \simeq \overline{\varepsilon}_{02}$.
Then, the ratio of the diffraction coefficients is 
approximately $\sigma = 1/2$, and we use this value in the numerical 
simulations presented below.

For spatially localized solutions,
the system (\ref{eq:chi2}) conserves the Hamiltonian
\begin{eqnarray*}
 H & = & \int_{-\infty}^{+\infty}{ \left\{ 
     {\left| \frac{\partial u}{\partial x} \right|}^2 + 
     \frac{\sigma}{2} {\left| \frac{\partial v}{\partial x} \right|}^2
     - \nu_1(x) {\left| u \right|}^2
     - \frac{\nu_2(x)}{2} {\left| v \right|}^2 \right.} \\
& &  {\left. -  \sum_n \delta \left( x-x_n \right)
                         {\left[  \beta_1 {\left| u \right|}^2
            + \frac{\beta_2}{2} {\left| v \right|}^2
            + {\rm Re} {\left( u^2 v^* \right)} \right]}
    \right\} dx} 
\end{eqnarray*}
and the total power 
$P=\int_{-\infty}^{+\infty} \left( |u|^2+|v|^2 \right) dx$.

\section{Periodic layered structures} \label{sect:period}
\subsection{General formalism} \label{sect:period_general}

\begin{figure}
\setlength{\epsfxsize}{11.5cm}
\vspace*{-10mm}
\hspace*{-0.3cm}\centerline{\mbox{\epsfbox{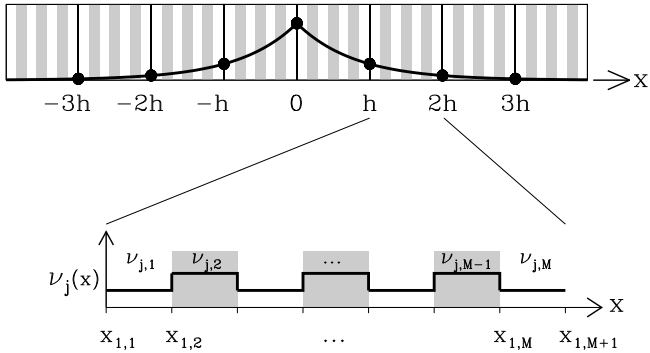}}}
\vspace*{-8mm}
\caption{ \label{fig:bragg}
The slab waveguide from Fig.~\ref{fig:model} viewed from the front with
indication of the substructure and notation of the identical linear
media (photonic crystal with $M$ sub-layers with different refractive
indices) in between the $\chi^{(2)}$ interfaces. The separation
between the nonlinear interfaces is $h$. A localized mode centered
at $x$=0 is also illustrated.
}
\end{figure}

To develop a general formalism for describing stationary, spatially
localized modes, we consider an infinite system of uniformly spaced nonlinear interfaces located periodically at the positions \mbox{$x_n = n h$}, which separate identical linear layers, \mbox{$\nu_{j} \left( x + n h \right) \equiv \nu_{j} \left( x \right)$}. Example of such a structure is shown in Fig.~\ref{fig:bragg}. This type of a {\em one-dimensional (1D) nonlinear photonic crystal} (NPC) can be used to prohibit light propagation along the transverse $x$ axis under certain conditions, resulting in field localization, and it resembles the operation of the so-called photonic crystal fiber in the linear regime~\cite{fiber}. The fundamental properties of NPC can be understood by studying {\em nonlinear localized modes}. Such modes appear in the frequency gap of the layered linear structure, and they can be called {\em discrete gap solitons} or, in a broader context, {\em intrinsic localized modes}~\cite{braun}.

\begin{figure}
\setlength{\epsfxsize}{8cm}
\centerline{\mbox{\epsfbox{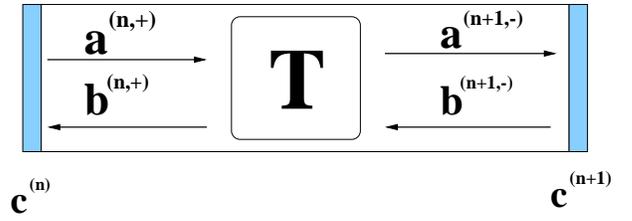}}}
\vspace{4mm}
\caption{ \label{fig:slice}
Transfer matrix relating the amplitudes at the boundaries of a linear
layer, independently for the FF and SH components,
see Eq.~(\ref{eq:lab}).}
\end{figure}

We search for solutions of Eqs.~(\ref{eq:chi2}) in the form:
\[
      u \left( x, z \right) = c_1 \left( x \right) e^{i \lambda_1 z} , 
           \quad
      v \left( x, z \right) = c_2 \left( x \right) e^{i \lambda_2 z} , 
\]
where $c_j \left( x \right)$ describe the transverse profiles of FF and SH
waves, respectively, and the real propagation constants $\lambda_j$ satisfy the phase-matching condition $\lambda_2 = 2 \lambda_1$.
Then, as the FF and SH fields do not interact in a linear bulk medium, each of them can be presented as a composition of two linear eigenmodes, corresponding to pairs of counter-propagating waves with opposite wavevector components along the $x$ axis. Regardless the internal structure of a linear layer between nonlinear interfaces, the wave amplitudes at its boundaries can be related by a transfer matrix:
\begin{equation} \label{eq:lab}
  \left( \begin{array}{l} a_j^{(n+1,-)} \\ 
         b_j^{(n+1,-)} \end{array} \right)
  = T^{(j)} \left( \lambda_j \right)
  \left( \begin{array}{l} a_j^{(n,+)} \\ 
                 b_j^{(n,+)} \end{array} \right).
\end{equation}
Here indices $\pm$ stand for the waves on the right and on the left of a given nonlinear layer, respectively, and the amplitudes of counter-propagating waves are denoted by $a_j$ and $b_j$, as shown in Fig.~\ref{fig:slice}.

In order to calculate the dependence of the matrix elements on the propagation constants, we should solve the corresponding linear problem. For practical applications, NPC can be produced by embedding nonlinear layers in an otherwise linear Bragg grating structure, see, e.g., Ref.~\cite{vilaseca2}. Thus, we assume that each linear layer consists of several sub-layers with a constant refractive index, see Fig.~\ref{fig:bragg}, i.e. $\nu_j(x) = \nu_{j,m}$ for $x_{n,m} \le x \le x_{n,m+1}$, where $m$ is used to number the sub-layer inside a linear slice ($1 \leq m \leq M$), and we have, by definition, $x_{n,1} = x_{n-1,M+1} = x_{n} = n h$. Then, the field in each of the sub-layers can be written as:
\begin{equation} \label{eq:cj}
  c_j \left( x \right) = 
     a_j^{(n,m)} e^{- {\mu}_{j,m} \left( x - x_{n,m} \right)}
     +
     b_j^{(n,m)} e^{{\mu}_{j,m} \left( x - x_{n,m} \right)} . 
\end{equation}
By definition, the amplitudes from 
Eq.~(\ref{eq:lab}) are $a_j^{(n,+)} = a_j^{(n,1)}$, 
$a_j^{(n+1,-)} = a_j^{(n,M+1)}$, with similar relations holding for
$b_j^{(n,\pm)}$.

Transverse wavenumbers can be calculated using the dispersion relations, $\lambda_1 = {\mu}_{1,m}^2 + \nu_{1,m}$ and $\lambda_2 = \sigma {\mu}_{2,m}^2 + \nu_{2,m}$, together with the phase matching condition $\lambda_2 = 2 \lambda_1$. From these expressions we conclude that, in general, for given $\lambda_j$, the waves can be either {\em localized} (i.e., ${\mu}_{j,m}$ is real) or {\em propagating} (i.e., ${\mu}_{j,m}$ is imaginary). 
Then, we notice that the waves in a multi-layered linear medium can be determined as {\em linear eigenmodes localized at the sub-layer boundaries}. 
First, from Eq.~(\ref{eq:cj}) it follows that the variation of the field amplitude through the $m$-th sub-layer of the width $h_m \equiv x_{n,m+1} - x_{n,m}$
is characterized by the following transfer matrix:
\begin{equation} \label{eq:T_p}
   T_{\rm p}^{(j,m)} = \left( \begin{array}{ll}
                                e^{ - \mu_{j,m} h_m}, & 0 \\
                                0,                    & e^{ \mu_{j,m} h_m}
                       \end{array} \right) .
\end{equation}
Second, the variation of the wave amplitudes at the boundary can be calculated by applying the field continuity conditions following from the model~(\ref{eq:chi2}). Specifically, we equate the amplitudes $c_j$ and their derivatives $d c_j / d x$ on both sides of the interface, and find the transfer matrix accounting for the field localization at a boundary separating the \mbox{$m$-th} and \mbox{$(m+1)$-th} sub-layers, 
\begin{equation} \label{eq:T_r}
 T_{\rm r}^{(j,m)} = \frac{1}{2} 
      \left( \begin{array}{ll}
        {1} + t_{\rm r}^{j,m}, &
        {1} - t_{\rm r}^{j,m} \\
        {1} - t_{\rm r}^{j,m}, &
        {1} + t_{\rm r}^{j,m}
    \end{array} \right) ,
\end{equation}
where $t_{\rm r}^{j,m} = \mu_{j,m} / \mu_{j,m+1}$. 
To simplify the further analysis, we include all linear properties in the transfer matrix.
Then, $\mu_{j,M+1} \equiv \mu_{j,1}$ due to periodicity of the underlying grating, and the linear $\delta$-response is characterized by the following matrix:
\begin{equation} \label{eq:T_delta}
 T_{\rm \delta}^{(j,m)} = \frac{1}{2} 
      \left( \begin{array}{ll}
        {2} + t_{\rm \delta}^{j,m}, &
        t_{\rm \delta}^{j,m} \\
        - t_{\rm \delta}^{j,m}, &
        {2} - t_{\rm \delta}^{j,m}
    \end{array} \right) ,
\end{equation}
where $t_{\rm \delta}^{j,m} = \beta_{j} / \mu_{j,m}$, and $m$ is the index of the linear layer with the delta-interface (in our case, $M+1$).
The total transfer matrix can be then found as a product,
\begin{equation} \label{eq:T_j}
 T^{(j)} = T_{\rm \delta}^{(j,M+1)} 
         T_{\rm r}^{(j,M)} T_{\rm p}^{(j,M)} \ldots
          T_{\rm r}^{(j,1)} T_{\rm p}^{(j,1)} .
\end{equation}

After calculating the matrix elements, we express the field in terms of the mode amplitudes defined at the nonlinear interfaces, $c_j^{(n)} = c_j \left( x_n \right)$, combining the relation~(\ref{eq:lab}) with the continuity conditions at the layers, $c_j^{(n)} = a_j^{(n,1)} + b_j^{(n,1)} = a_j^{(n,M+1)} + b_j^{(n,M+1)}$. Integrating Eqs.~(\ref{eq:chi2}) over small intervals including the nonlinear interfaces at the positions $x_n$ (note that we exclude linear responses as they are already accounted for in the transfer matrices), we derive the so-called {\em discrete  $\chi^{(2)}$ equations}:
\begin{equation} \label{eq:dchi2}
 {\displaystyle \begin{array}{l}
    \eta_1 U_n + \left( U_{n-1} + U_{n+1} \right)
              + \chi_1 U_n^* V_n = 0 ,
               \\*[9pt]
    \eta_2 V_n + \left( V_{n-1} + V_{n+1} \right)
              + \chi_2 U_n^2 = 0 ,
  \end{array} } 
\end{equation}
where $U_n=c_1^{(n)} \sqrt{ |\xi_1 \xi_2| }$ and $V_n = c_2^{(n)} |\xi_1|$ 
are the normalized FF and SH amplitudes at the $n$-th 
nonlinear layer. Parameters $\chi_j$ and $\eta_j$ are defined by 
the matrix elements,
\begin{equation} \label{eq:dparam}
 \begin{array}{l} 
  {\displaystyle 
    \xi_1 =  - T_s^{(1)} / {\left( 2 \mu_{1,1} \right)}, \;
    \xi_2 =  - T_s^{(2)} / {\left( 2 \sigma \mu_{2,1} \right)}
  } \\*[9pt] {\displaystyle 
    \eta_j = - {\rm Tr}( T^{(j)} ),  \;  
    \chi_j = {\rm sign} {\left( \xi_j \right)}, 
  } \end{array}
\end{equation}
where
\begin{equation} \label{eq:T_char}
 \begin{array}{l} 
  {\displaystyle 
    T_s^{(j)} = T_{[1,1]}^{(j)} + T_{[2,1]}^{(j)} 
                - T_{[1,2]}^{(j)} - T_{[2,2]}^{(j)},
  } \\*[9pt] {\displaystyle 
    {\rm Tr}( T^{(j)} ) \equiv 
                T_{[1,1]}^{(j)} + T_{[2,2]}^{(j)} .
  } \end{array}
\end{equation}
Hereafter we use the notation $T_{[n,m]}^{(j)}$ to denote the matrix element in the row $n$ and column $m$ of the matrix $T^{(j)}$. It is easy to verify that the parameters $\eta_j$ and normalization coefficients $\xi_j$ are {\em real} for any (real) propagation constants $\lambda_j$. A proof of this fact, together with discussion of some other properties of the constructed transfer matrices, is given in Appendix~A.

We note that some particular cases of the system~(\ref{eq:dchi2}) have been earlier discussed (but, in fact, never derived in a consistent manner) in the analysis of the nonlinear interface dynamics under the condition of Fermi resonance~\cite{agran_grat,dubrov}, arrays of weakly interacting quadratic waveguides~\cite{bang,Lederer,miller,dom_wall}, and beam propagation in nonlinear lattices~\cite{cbc}.

\subsection{Discrete gap solitons} \label{sect:period_dgsol}

Now we use Eqs.~(\ref{eq:dchi2}) to find the stationary localized modes of nonlinear superlattices, or {\em discrete gap solitons}. A similar problem was earlier analyzed in Ref.~\cite{agran_grat} in the so-called continuum limit, when the modes become wide and are effectively supported by many interfaces, with the excitation profiles approaching those of quadratic solitons~\cite{solitons}. On the other hand, it has been demonstrated that discrete states in a closed system of few interfaces can have different topologies and possess quite peculiar properties~\cite{dubrov,bang}. For the case at hand, when the number of interfaces is infinite (e.g. much larger than the characteristic mode width), the different types of highly localized waves have been identified~\cite{Lederer}, and their profiles were described by approximate analytical solutions. However, until now the transitional case of moderately localized modes has not been addressed. Thus, we develop a more complete analytical description of discrete gap solitons, which can predict their properties in all the parameter regions.

In order to find approximate solutions for highly localized modes, we use the variational method. First, we have to choose the trial functions. We use the fact that in the high localization limit the tails are almost linear, so that the amplitudes decay nearly exponentially. Then, we introduce two sets of trial functions to account for different topologies~\cite{Lederer}: {\em odd modes}, when the center of symmetry is located at a layer,
\begin{equation} \label{eq:testf_odd}
    U_n^{\rm (o)} = U_0 s_1^{|n|} e^{- \rho_1 |n| }, \; 
    V_n^{\rm (o)} = V_0 s_2^{|n|} e^{- \rho_2 |n| }, 
\end{equation}
and {\em even modes}, when the center of symmetry is located between two neighboring layers,
\begin{equation} \label{eq:testf_even}
 {\displaystyle \begin{array}{l}
    U_n^{\rm (e)} = 
      {\left\{ \begin{array}{l}
        U_0 s_1^{|n|} e^{- \rho_1 |n| }, \; n \ge 0,\\
        U_0 t s_1^{|n+1|} e^{- \rho_1 |n+1| }, \; n < 0,
        \end{array} \right.}
          \\*[9pt]
    V_n^{\rm (e)} = 
      {\left\{ \begin{array}{l}
        V_0 s_2^{|n|} e^{- \rho_2 |n| },  \; n \ge 0,\\
        V_0 s_2^{|n+1|} e^{- \rho_2 |n+1| }, \; n < 0.
        \end{array} \right.}
  \end{array} }
\end{equation}
Here the parameters $s_j = \pm 1$ are introduced to describe {\em unstaggered} and {\em staggered} profiles, and $t = \pm 1$ to produce either {\em untwisted} or {\em twisted} modes (for the signs "$+$" or "$-$", respectively).

After selecting the mode topology and fixing the values of $s_j$ and $t$, the unknown values $U_0$, $V_0$, $\rho_j$ ($\rho_j>0$) are determined by minimizing the Lagrangian corresponding to Eqs.~(\ref{eq:dchi2}).
Details of these calculations will be presented elsewhere~\cite{malomed}, here we give a brief summary of the main results, and discuss their physical consequences.

As the system (\ref{eq:dchi2}) possesses the symmetries (i)~$\chi_1 \rightarrow - \chi_1$, $s_1 \rightarrow - s_1$, $\eta_1 \rightarrow - \eta_1$, and (ii)~$\chi_2 \rightarrow - \chi_2$ and $V_n \rightarrow - V_n$, we consider, without a lack of generality, the case $\chi_j = 1$. The analysis shows that localized solutions exist only if $s_1 \eta_1 < -2$. The latter condition means that the FF component is {\em unstaggered} for $\eta_1 < -2$, and {\em staggered}, otherwise ($\eta_1 > 2$). Similarly, we consider only the case $|\eta_2| > 2$, as for other values the localized solutions are unstable due to resonant interaction with linear waves~\cite{Lederer}. 

\begin{figure}
\setlength{\epsfxsize}{8cm}
\centerline{\mbox{\epsfbox{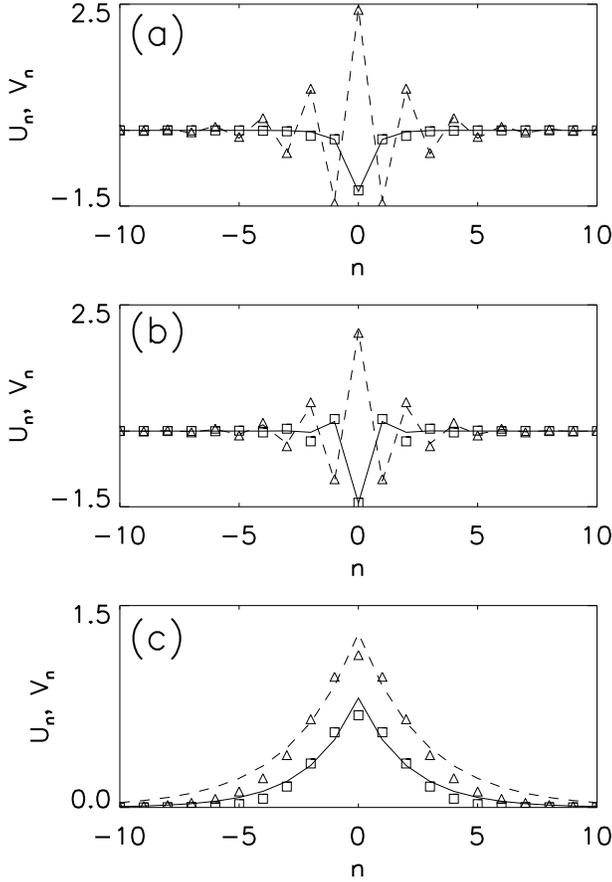}}}
\vspace*{5mm}
\caption{ \label{fig:odd}
Different types of odd two-frequency localized modes 
(numerical solutions: FF -- triangles, SH -- squares; 
 variational profiles: FF -- dashed line, SH -- solid).
(a)~staggered FF and unstaggered SH ($\eta_1 = 2.4$, $\eta_2 = 4.5$), 
(b)~both components staggered ($\eta_1 = 2.4$, $\eta_2 = 3$), 
(c)~both unstaggered ($\eta_1 = -2.4$, $\eta_2 = -3.5$). }
\end{figure}

Analyzing the linear problem, it is straightforward to see that the SH mode can be staggered only if $\eta_2 > 2$. Then, we distinguish between two limits: (i)~a strongly localized FF mode ($\eta_1 \gg 2$), when the SH consists of staggered linear tails~\cite{Lederer}, and (ii)~the cascading limit ($\eta_2 \gg 2$), when the SH profile is unstaggered and can be found as $V_n \simeq U_n^2 / \eta_2$, resulting in an effectively cubic nonlinearity for the FF wave~\cite{miller} (see also~\cite{solitons}). In the intermediate case, a transition of the SH profile between staggered and unstaggered topologies should be observed. Indeed, our variational calculations predict that the SH is staggered for $2 < \eta_2 < \eta_{22}$, and unstaggered for $\eta_2 > \eta_{22}$. Here, the critical parameter value depends on $\eta_1$: for odd modes, it is found from the quadratic equation, $\sqrt{\eta_{22}} + 1 / \sqrt{\eta_{22}} = |\eta_1|$, and a similar relation holds for even modes $\sqrt{\eta_{22}+1} + 1 / \sqrt{\eta_{22}+1} = |\eta_1|$. The topology does not change sharply as the parameter $\eta_2$ crosses the critical value $\eta_{22} \left( \eta_1 \right)$, but numerical calculations confirm that such a transition occurs in a region close to the separation line.

\begin{figure}
\setlength{\epsfxsize}{8cm}
\centerline{\mbox{\epsfbox{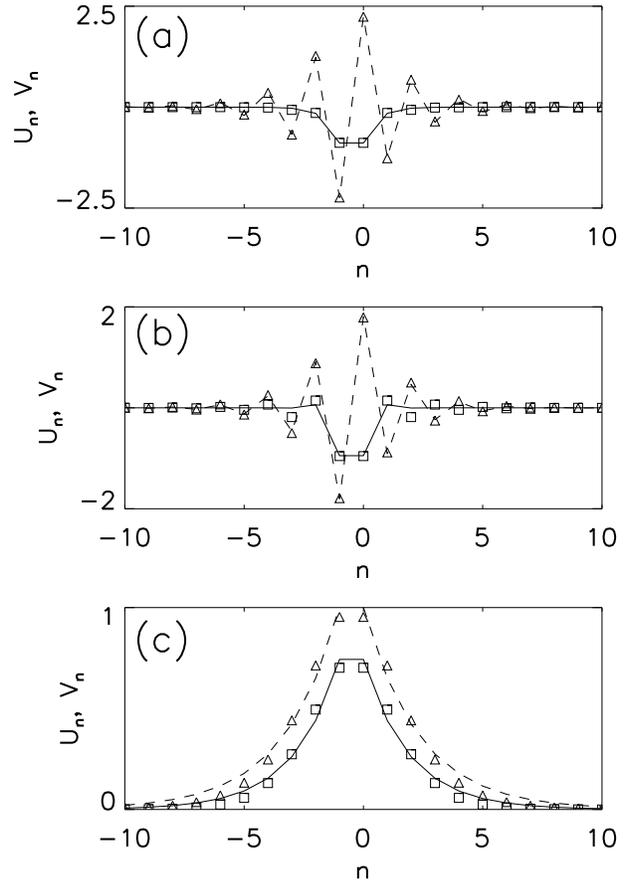}}}
\vspace*{5mm}
\caption{ \label{fig:even}
Characteristic examples of even two-frequency localized modes 
with untwisted FF component; 
the notations are the same as in Fig.~\ref{fig:odd}. 
Parameter values are 
(a)~$\eta_1 = 2.45$, $\eta_2 = 4.5$,
(b)~$\eta_1 = 2.45$, $\eta_2 = 2.5$,
(c)~$\eta_1 = -2.45$, $\eta_2 = -3$. }
\end{figure}

We performed numerical analysis and found that the approximate variational solutions provide close matching for the highly localized profiles, i.e. for relatively large values of $|\eta_j|$. Examples of odd and even modes are presented in Figs.~\ref{fig:odd} and \ref{fig:even}, respectively. We see that the profiles of staggered modes, supported by only a few interfaces, are described very accurately. However, for wider modes the deviations between the exact numerical and variational solutions are more pronounced, see Figs.~\ref{fig:odd}(c) and \ref{fig:even}(c). The limitation of the variational solution is due to the specific choice of the trial functions~(\ref{eq:testf_odd}),(\ref{eq:testf_even}), which are not suitable for description of ``moderately'' localized waves with smoother profiles.

It can be demonstrated that in the continuum limit \mbox{($\eta_j \rightarrow -2^{-}$)} untwisted modes acquire the profiles of quadratic solitons~\cite{solitons}, that can be well approximated with the ${\rm sech}$-type functions~\cite{me}. A special {\em quasi-continuous approach}, which allows to determine the mode profiles as a soliton bound state can also be developed. The resulting approximate solutions provide very good estimates, and they are very useful for understanding the mode scaling properties, i.e. a change from broad solitons to narrow highly localized states. On the other hand, it should be noted that the twisted modes do not exist close to the continuum limit, because their profiles are intrinsically discrete due to a sharp amplitude change between the layers at the mode center. A comprehensive description of these results goes beyond the scope of the present paper, and will be presented elsewhere~\cite{malomed}.

\section{ A single $\chi^{(2)}$ nonlinear layer 
          embedded in a periodic structure }    \label{sect:single}

Let us now consider a special situation, when there exists only a single nonlinear layer located at \mbox{$x_0 = 0$} embedded into a linear grating. A similar problem has been considered in our recent paper~\cite{PRE} for the case when the linear media on both sides of the nonlinear interface are uniform. Here, we generalize those results to the case of non-uniform linear media, considering a nonlinear layer with quadratic nonlinearity embedded into a periodic structure. Therefore, we modify the model~(\ref{eq:chi2}) 
as follows: (i)~nonlinear coupling terms are taken into account only for $n=0$, and (ii)~a linear response of the central layer is assumed to be different from that of other layers, i.e. we change $\beta_j \rightarrow \alpha_j + \beta_j$ at $n=0$.
In connection with the previous problem (see Secs.~\ref{sect:model} and~\ref{sect:period}), the case with $\alpha_j=0$ corresponds to the limit of a highly localized mode, when the mode width is much smaller than the distance in between the nonlinear layers (see Fig.~\ref{fig:bragg}).

\begin{figure}
\vspace*{-3.4cm}
\setlength{\epsfxsize}{8cm}
\centerline{\mbox{\epsfbox{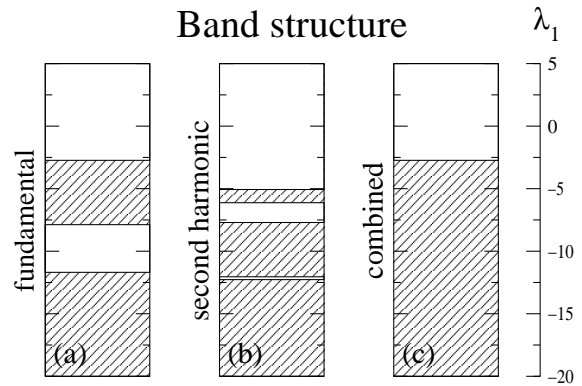}}}
\vspace*{-2.7cm}
\caption{ \label{fig:lgrat}
The band structure for (a)~FF and (b)~SH modes, and (c)~combined band gap (additional narrow gaps in FF and SH spectra exist for smaller $\lambda_1$, not shown). Areas without shading correspond to band gaps.
Parameters are $h_{\pm} = 0.6$, $\nu_{1,+} = 0$, $\nu_{1,-} = -6$, $\nu_{2,+} = -8$, $\nu_{2,-} = -13$, and $\beta_j^{\pm}=0$.
}
\end{figure}

Following the general approach outlined in the previous section, we first analyze the linear properties. We also use the similar notations, but for a single nonlinear interface we omit the index $n=0$. From the theory of linear structures, it follows that a link between the linear-wave amplitudes can be characterized by the reflection coefficients, $r_j^+ = b_j^+ / a_j^+$ and $r_j^- = a_j^- / b_j^-$. We do not assume that the linear structure is symmetric [in general $\nu_j(x) \ne \nu_j(-x)$] and denote with $+$ and $-$ the wave characteristics at the right and left boundaries of the nonlinear layer. If the linear structure is periodic, the coefficients $r_j^{\pm}$ can be found by solving the following eigenvalue problems:
\begin{equation} \label{eq:r_eigen}
      T^{(j,\pm)}
      \left( \begin{array}{l} 1 \\ r_j^{\pm} \end{array} \right) 
    =
      \tau^{(j,\pm)} 
      \left( \begin{array}{l} 1 \\ r_j^{\pm} \end{array} \right) .
\end{equation}
Here $T^{(j,\pm)}$ is the transfer matrix of one linear segment in a periodic lattice, starting on the right ($+$) or the left ($-$) side of a nonlinear interface, and it can be calculated using \mbox{Eqs.~(\ref{eq:T_p})-(\ref{eq:T_j})}. 
Then, we notice that the determinants are: $||T_{\rm p}^{(j,m,\pm)}|| = 1$, $||T_{\rm \delta}^{(j,m,\pm)}|| = 1$, $||T_{\rm r}^{(j,m,\pm)}|| = \mu_{j,m}^{\pm} / \mu_{j,m+1}^{\pm}$, and thus $||T^{(j,\pm)}|| \equiv 1$. Using this relation, we solve Eq.~(\ref{eq:r_eigen}) and find the values of the reflection coefficients:
\begin{equation} \label{eq:r_j}
  r_j^{\pm} = T_{[2,1]}^{(j,\pm)}  
           {\left(  \tau^{(j,\pm)} - T_{[2,2]}^{(j,\pm)} \right)}^{-1} ,
\end{equation}
where 
\[
     \tau^{(j,\pm)} = - (\eta_j^{\pm} / 2) \left[ 1 - 
           \sqrt{ 1 - {\left(2 / \eta_j^{\pm}\right)}^2 } \right]. 
\]
It can be proved that $\eta_j^{\pm}$ are real (see Appendix~A). Then, we see that the modes decay at infinity, i.e. $|\tau^{(j,\pm)}|<1$, only if $|\eta_j^{\pm}| > 2$. This condition, which has already been mentioned in Sec.~\ref{sect:period}, defines a {\em band gap}, when the wave propagation is prohibited for a specific range of the propagation constant $\lambda_j$ (see an example in Fig.~\ref{fig:lgrat}). 

By applying the continuity conditions, $c_j = a_j^\pm + b_j^\pm$,
we obtain the expressions for the amplitudes at the nonlinear layer:
$|c_1|^2 = \widetilde{\alpha}_1 \widetilde{\alpha}_2$,
$c_2 = - \widetilde{\alpha}_1$,
where
$\widetilde{\alpha}_1 = \alpha_1 - {\zeta_1^+} - {\zeta_1^-}$,
$\widetilde{\alpha}_2 = \alpha_2 - \sigma {\zeta_2^+} - \sigma {\zeta_2^-}$,
and ${\zeta_j^\pm} = \mu_j^\pm {\left( 1 - r_j^\pm \right)} 
                     / {\left( 1 + r_j^\pm \right)}$ are the effective
transverse wave numbers at the right ($+$) and left ($-$) boundaries of the nonlinear layer; their values are determined by the dispersion relation of the periodic linear gratings. As has been demonstrated in the previous section, the wave numbers $\mu_j^\pm$, the linear transfer matrices, and, according to Eq.~(\ref{eq:r_j}), the reflection coefficients $r_j^\pm$ depend on the propagation constants $\lambda_j$, which, in turn, are related by the phase-matching condition $\lambda_2 = 2 \lambda_1$. Thus, for fixed physical characteristics, the localized modes constitute {\em a one-parameter family}, and we choose $\lambda_1$ as a free parameter. We note that the coefficients $\zeta_1^{\pm}$ are real in the band gap
(see Appendix~B).

\begin{figure}
\setlength{\epsfxsize}{10cm}
\centerline{\mbox{\epsfbox{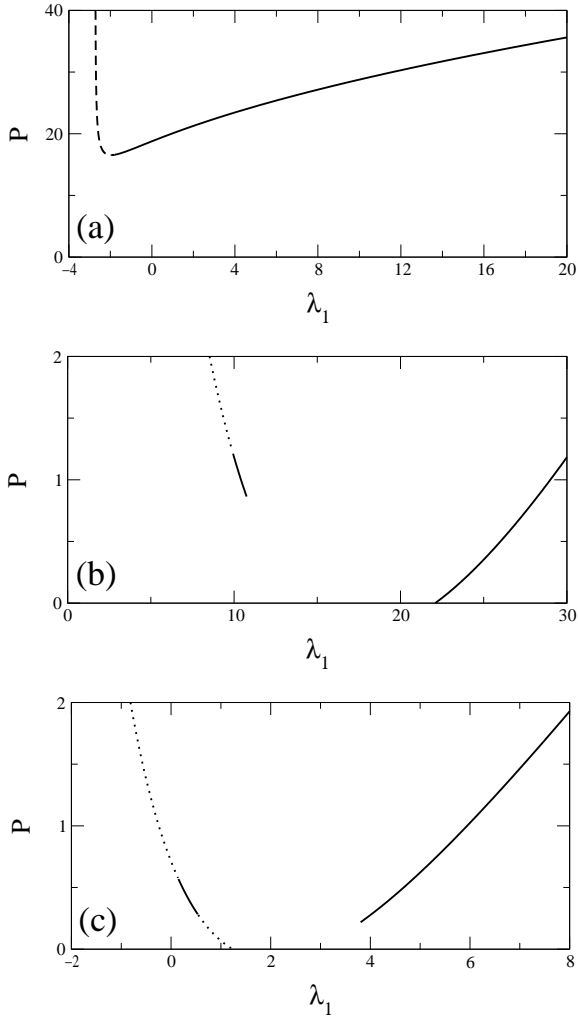}}}
\vspace*{5mm}
\caption{ \label{fig:sPH}
Three types of the power dependences for two-color stationary localized modes, for different linear mismatches:
(a)~$\alpha_1 = -1$ and $\alpha_2 = -1$; 
(b)~$\alpha_1 = 10$ and $\alpha_2 = 8$; 
(c)~$\alpha_1 = 4$ and $\alpha_2 = 6$. 
Solid~-- stable, dashed~-- unstable, and dotted~-- oscillatory unstable.
Parameters of the linear structure correspond to those in Fig.~\ref{fig:lgrat}. }
\end{figure}

In order to illustrate the features of nonlinear modes, we consider a structure similar to that used in experiments on the SH generation in localized modes~\cite{vilaseca}. In the experimental setup, a periodic linear grating was built of two materials with different refractive indices, characterized by the parameters $\nu_{j,\pm}$, with corresponding finite widths of the layers, $h_{\pm}$. The nonlinear interface was created by cutting the grating in two parts and coating the interface to enhance the effective quadratic nonlinearity. Characteristics of the defect layer were controlled by adjusting the gap. 
An example of the band gap for such a photonic structure is presented in Fig.~\ref{fig:lgrat}. Note that because the linear structures on either sides of the nonlinear layer are chosen to be the same [up to a constant shift, in our notation $\nu_{j,-}(x) = \nu_{j,+}(x + h_{+})$], the corresponding gaps coincide.

Many of the properties of the localized states can be understood by analyzing the power diagram $P \left( \lambda_1 \right)$. Characteristic examples of such a dependence are shown in Figs.~\ref{fig:sPH}(a,b,c). Similar to the case of homogeneous linear media~\cite{PRE}, there always exists a branch in a parameter region unbounded from above for $\lambda_1$ larger than some critical value. Quite remarkably, the corresponding mode properties are very similar to those of {\em quadratic solitons}~\cite{solitons}. In particular, for large values of the propagation constant $\lambda_1$ the power dependence $P \left( \lambda_1 \right)$ always has a positive slope, while for smaller $\lambda_1$ the slope can be negative, resulting in {\em bistability}. Such a case is demonstrated in Fig.~\ref{fig:sPH}(a). We found that stability of the corresponding modes can be determined by the Vakhitov-Kolokolov criterion, i.e. the localized modes are stable provided $d P / d \lambda_1 >0$, and unstable, otherwise.

\begin{figure}
\setlength{\epsfxsize}{8cm}
\centerline{\mbox{\epsfbox{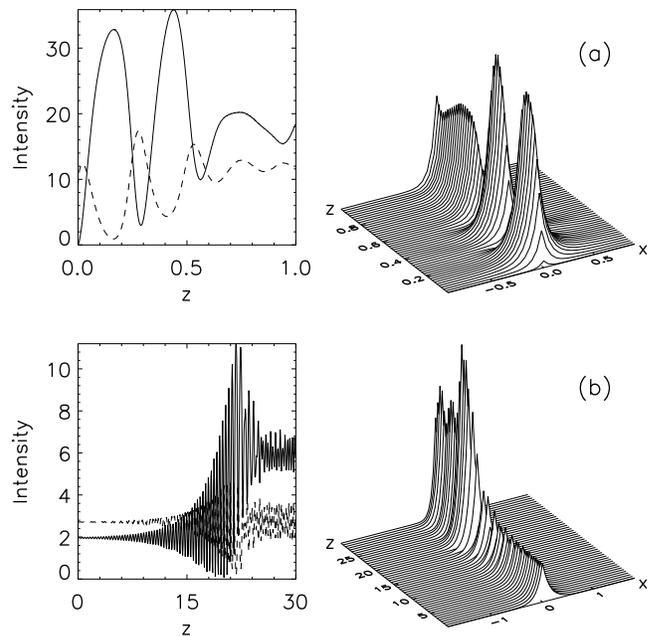}}}
\vspace*{5mm}
\caption{ \label{fig:shg}
(a)~Generation of a stable localized mode from an input FF beam with a Gaussian transverse profile (initial power \mbox{$P \simeq 8$});
(b)~Switching from a perturbed unstable (corresponding to $\lambda_1=-1$)
to a stable mode. 
Left:~the field intensities of the FF (dashed) and SH (solid) components at the interface. 
Right:~evolution of the SH field.
Parameters correspond to those in Fig.~\ref{fig:sPH}(c).}
\end{figure}

On the other hand, the spectrum of a linear periodic structure consists of several bands. Moreover, even inside a band gap the modes cannot exist if the condition $\widetilde{\alpha}_1 \widetilde{\alpha}_2>0$ is not satisfied. Thus, in a sharp contrast with two-color parametric solitons in homogeneous media, other branches can appear for smaller values of $\lambda_1$, and we found that the localized mode properties can be very different compared to the modes corresponding to the right branch. For example, for Figs.~\ref{fig:sPH}(b,c), the left branches at higher intensities correspond to smaller $\lambda_1$ and wider profiles. This happens because the SH amplitude is negative, $c_2<0$, which results in  effectively {\em self-defocusing} nonlinear response. It is interesting to note that similar types of power dependencies occur for a layer possessing a self-defocusing Kerr-type nonlinearity~\cite{nls_imp}, and the corresponding localized modes are stable. We have performed a linear stability analysis and have found that, similar to the case of the Kerr-type nonlinearity, the Vakhitov-Kolokolov type criterion {\em cannot be applied to determine stability} of such modes existing in an effectively defocusing medium. However, our calculations demonstrate that {\em parametric resonances can lead to oscillatory instability}, as illustrated in Figs.~\ref{fig:sPH}(b,c). A comprehensive discussion of the mode stability will be presented elsewhere.

The two-color modes can be generated by launching a localized FF wave at the interface, as shown in Fig.~\ref{fig:shg}(a). We have also studied the evolution of perturbed unstable modes, which can evolve toward a stable state. Example in Fig.~\ref{fig:shg}(b) demonstrates the development of an oscillatory instability with subsequent switching to a stable branch.

\section{Conclusions}

We have developed a general formalism for analyzing spatially localized nonlinear modes (discrete gap solitons) in periodic photonic structures with embedded quadratic (or $\chi^{(2)}$) nonlinear interfaces~--- {\em nonlinear quadratic superlattices}. Our approach can be applied to different types of periodic linear media with isolated or periodic nonlinear interfaces, where nonlinearity can support parametric interaction and generation of the second-harmonic field. In the case of a nonlinear superlattice, i.e. periodically spaced layers with quadratic nonlinear response, we have derived an effective discrete model. For such a periodic structure, we have found discrete gap solitons of different topologies in the form of fundamental and second-harmonic fields coupled parametrically at the nonlinear interfaces.

For a single nonlinear layer embedded in a linear periodic medium, we have described a novel type of nonlinear localized defect mode~--- a two-color photonic mode. Some of the properties of these two-color localized modes, such as stability, generation, and switching, have been shown to be remarkably similar with those of quadratic parametric solitons in homogeneous media. However, we have revealed a number of specific properties of such modes, and also demonstrated their generation from a localized beam of the fundamental frequency, as well as switching from an unstable to a stable state.

We believe our results are important, on one hand, for the theory of nonlinear photonic crystals where nonlinearities appear due to phase-matched harmonic generation, and, on the other hand, for creating tunable band gap materials where gaps could be opened or closed depending on the input intensity. 

\section*{Acknowledgments}

The authors are indebted to F. Lederer, N.~N. Rosanov, and R. Vilaseca for useful comments. Costas Soukoulis acknowledges the hospitality of the Optical Sciences Centre.
The work has been partially supported by the Planning and Performance Fund of the Institute of Advanced Studies, by the Australia-Denmark collaborative project of the Department of Industry, Science, and Tourism, by the Australian Photonics Cooperative Research Centre, and by  the Danish Technical Research Council under Talent Grant No.~9800400.

\section*{Appendix A: Transfer matrix properties}

In order to demonstrate some features of linear modes existing in periodic structures, we consider the properties of the corresponding transfer matrices. First, we see that a matrix describing changes of the wave amplitudes at the boundary between linear layers, $T_{\rm r}$, depends only on the wave numbers on either side of the interface, as follows from Eq.~(\ref{eq:T_r}). It is also easy to check that the following relation holds:
\begin{equation} \label{eq:Tr_mut}
   \begin{array}{l} {\displaystyle 
      T_{\rm r}^{(j,m)} \equiv T_{\rm r}(\mu_{j,m+1}, \mu_{j,m}) =
   } \\*[9pt] {\displaystyle \qquad \qquad
      T_{\rm r}(\mu_{j,m+1},\tilde{\mu}) \;
      T_{\rm r}(\tilde{\mu},\mu_{j,m}) ,
   } \end{array}             
\end{equation}
where $\tilde{\mu}$ is arbitrary, and we assume that it is real. Let us introduce a new matrix $\widetilde{T}^{(j)} = T_{\rm r}(\tilde{\mu},\mu_{j,M+1}) T^{(j)} T_{\rm r}(\mu_{j,1},\tilde{\mu})$, and use Eq.~(\ref{eq:Tr_mut}) to present it in a special form:
\[
   \widetilde{T}^{(j)} = \widetilde{T}_{\rm \delta}^{(j,M+1)} 
                         \widetilde{T}_{\rm p}^{(j,M)} \ldots
                         \widetilde{T}_{\rm p}^{(j,1)} .
\]
Here $\widetilde{T}_{\rm p}^{(j,n)} \equiv T_{\rm r}(\tilde{\mu},\mu_{j,n}) T_{\rm p}^{(j,n)} T_{\rm r}(\mu_{j,n},\tilde{\mu})$ and $\widetilde{T}_{\rm \delta}^{(j,n)} \equiv T_{\rm r}(\tilde{\mu},\mu_{j,n}) T_{\rm \delta}^{(j,n)} T_{\rm r}(\mu_{j,n},\tilde{\mu})$. It can be verified by direct substitution that these matrices are real (we use the fact that, for stationary modes, $\mu_{j,n}$ are real or purely imaginary), and therefore $\widetilde{T}^{(j)}$ is real as well.
Finally, we find that ${\rm Tr}( T^{(j)}) \equiv {\rm Tr}( \widetilde{T}^{(j)})$ and $T_s^{(j)} / \mu_{j,1} \equiv \widetilde{T}_s^{(j)} / \tilde{\mu}$, which proves that coefficients $\xi_j$ and $\eta_j$ in Eq.~(\ref{eq:dchi2}) are real.

\section*{Appendix B: Reflection coefficients for modes in a band gap}

Although it is possible to extend the technique presented in Appendix~A to prove the properties of the reflection coefficients in the case of an infinite linear grating (introduced in Sec.~\ref{sect:single}), here we employ a different approach. We note that for stationary waves in a band gap there should be no energy flow along the $x$ axis. The corresponding restrictions follows from Eq.~(\ref{eq:chi2}):
\begin{equation} \label{eq:eflow}
  {\rm Im}\left(u \frac{\partial u^{\ast}}{\partial x}\right) = 0, \quad
  {\rm Im}\left(v \frac{\partial v^{\ast}}{\partial x}\right) = 0,
\end{equation}
These conditions are satisfied when the amplitudes of the counter-propagating waves coincide, i.e. $|r_j^{\pm}| = 1$ if ${\mu}_j^{\pm}$ is imaginary. On the other hand, for the layers with real ${\mu}_j^{\pm}$ the linear modes should be in phase, i.e. ${\rm Im} (r_j^{\pm}) \equiv 0$. Then, it immediately follows that in a band gap the coefficients $\zeta_j^{\pm}$ are real.

\end{multicols}
\end{document}